\documentclass[twocolumn,showpacs,footinbib,10pt]{revtex4-1} %superscriptaddress
\usepackage{color}    
\usepackage[breaklinks]{hyperref}
\hypersetup{colorlinks=false}
\usepackage[hmargin=0.69in, vmargin=0.75in] {geometry}         %    % See geometry.pdf to learn the layout options. There are lots.

\usepackage{graphicx}
\usepackage{amssymb}
\usepackage{amsmath}
\usepackage{epstopdf}
\usepackage{rotating}
\usepackage{natbib}
\usepackage[section]{placeins}
\newcommand {\be}{\begin{equation}}
\newcommand {\ee}{\end{equation}}

\usepackage[caption=false, position=t,singlelinecheck=off]{subfig}

\begin{document}

\title{Entanglement dynamics 
 and Mollow nonuplets  between two coupled quantum dots in a nanowire photonic crystal system}
\author{Gerasimos Angelatos}
\email{g.angelatos@queensu.ca}
\author{Stephen Hughes}
\affiliation{Department of Physics, Engineering Physics and Astronomy, Queen's University, Kingston, Ontario, Canada K7L 3N6}
        % Activate to display a given date or no date
%\renewcommand{\abstractname}{Abstract}
\begin{abstract}
%\subsection*{Abstract}

We introduce a nanowire-based photonic crystal waveguide system capable of controllably mediating the photon coupling between two  quantum dots which are macroscopically separated.  Using a rigorous  Green-function-based master equation approach, our two-dot system is shown to provide  a wide range of interesting quantum regimes.  In particular, we demonstrate the formation of long-lived entangled states and study the resonance fluorescence spectrum which contains clear signatures of the coupled quantum dot pair. Depending upon the operating frequency, one can obtain a modified Mollow triplet spectrum or a Mollow nonuplet, namely  a spectrum with nine spectral peaks. These multiple peaks are explained in the context of photon-exchange-mediated dressed states.  Results are robust with respect to scattering loss, and  spatial filtering via propagation allows for each quantum dot's emission to be observed individually.

\end{abstract}
%, quantum light matter, quantum optical phenomina in amplifying media andcooperative phenomena, quantum dots, nanorods
\maketitle

\section{Introduction}  
The ability to mediate  coupling and entanglement between qubits  is important for optical quantum information systems~\cite{Ladd2010, Kimble2008}. In particular, it is  desirable that future quantum information systems are scalable, and should  operate {\it on-chip}, where, e.g., photons are manipulated in the plane of a waveguide. In addition,  the ability to produce and maintain entanglement between spatially separated qubits  is required, both for measurement purposes and to permit individual control of separated qubits.

Photonic crystal (PC) slabs~\cite{Yablonovitch1987, John1987, Johnson1999} with embedded quantum dots (QDs) are  strong candidates for on-chip quantum information systems~\cite{Yao2009,  Dalacu2010}, 
since they have the ability to  modify the local optical density of states (LDOS) through integrated cavities and waveguides.  Systems containing a single QD coupled to a PC antinode  can operate as a single photon source and facilitate the strong-coupling regime~\cite{Yao2009, Hennessy2007, BaHoang2012}.  However, semiconductor structures such as  PC slabs have yet to demonstrate coupling between multiple QDs in a controlled way.  This is largely due to the limitations of Stranski\textendash{}Krastanov growth, where the self-assembly of QDs results in limited control over their position and emission frequency, and  poor coupling to PC waveguide modes~\cite{Yao2009,  BaHoang2012}, such that coupling has so far only been demonstrated between QDs in a shared cavity~\cite{Laucht2010}.  Systems that couple QDs via an arbitrary length PC waveguide mode~\cite{Yao2009bell, Minkov2013} are desirable, offering the ability to excite and probe individual QDs. Coupling QDs via plasmonic waveguides has  been proposed~\cite{Gonzalez-Tudela2011}, though metallic systems suffer from  material losses and Ohmic heating.  These waveguide structures represent a rapid departure from a simple  Lorentzian cavity system, requiring a Green function approach to study the complex electromagnetic environments, i.e.,  with arbitrary losses and 
an inhomogeneous structure~\cite{Dung2002, Yao2009, Minkov2013}.

Photonic crystal structures comprised of arrays of dielectric rods~\cite{Johnson1999} offer an alternative to the traditional slab design~\cite{Johnson2001, Assefa2004}.  Moreover, semiconductor nanorod and nanowire (NW) fabrication techniques have seen  dramatic improvements in recent years~\cite{ Dubrovskii2009,  Makhonin2013} and the ability to produce QDs of deterministic position and optical properties in NWs has been demonstrated both during molecular-beam epitaxy growth~\cite{Tribu2008, Makhonin2013} and via post-process~\cite{Pattantyus-Abraham2009}.   Deterministic emitter placement has also been shown for nitrogen vacancy centers in diamond NWs~\cite{Babinec2010}.

\begin{figure}[h]
\centering
\vspace{-5pt}
\subfloat[]{\label{NWWG}\includegraphics[width=0.32\textwidth]{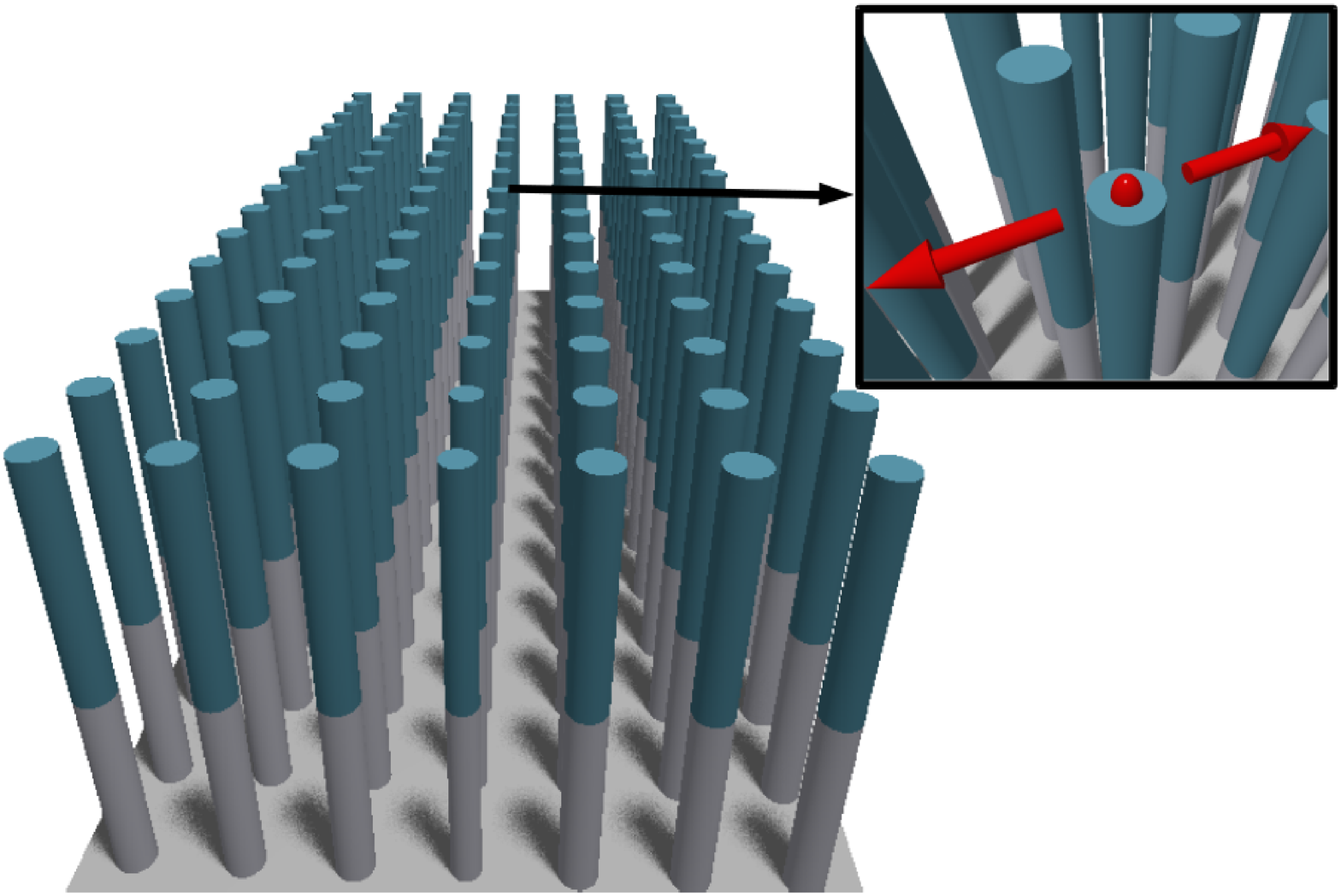}}
\subfloat[]{\label{Elev1}\hspace{2pt}\includegraphics[width=0.15\textwidth]{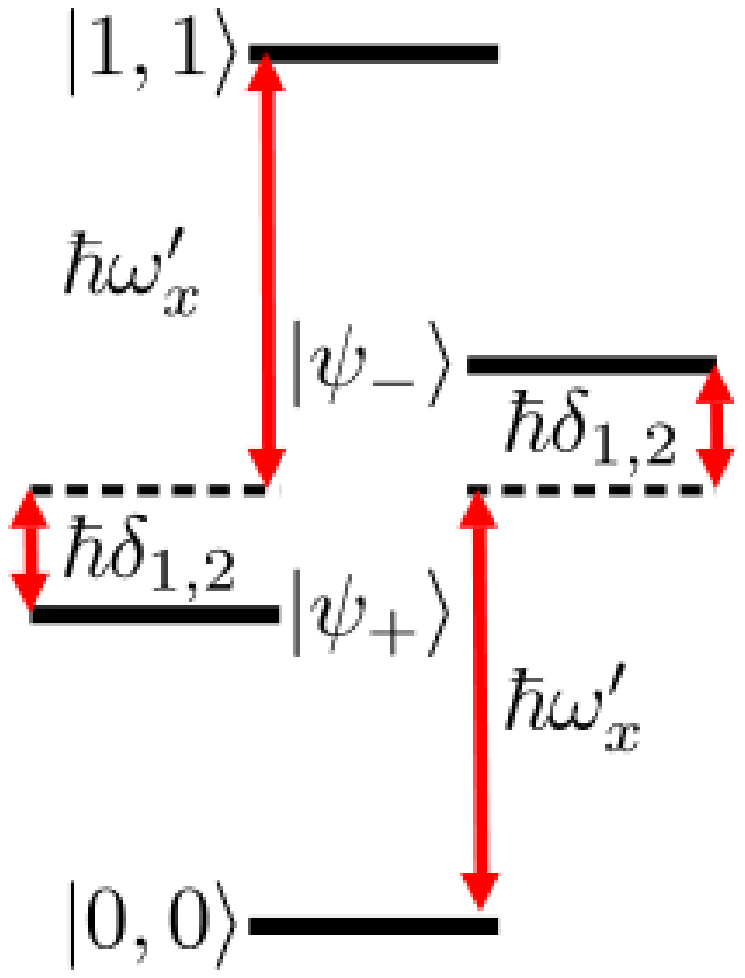}}
\vspace{-5pt}
\caption{ (Color online) (a) Proposed NW waveguide, which confines light to the central channel of reduced-radius NWs, with inset showing one of two QDs (size exaggerated) embedded in a NW and
  coupled with the other QD via the waveguide.  (b) Energy-level diagram (not to scale) of a two-QD system  interacting with the waveguide, with no drive and $\delta_{1,2}<0$, where $|\psi_{\pm}\rangle=\frac{1}{\sqrt{2}}(|0,1\rangle\pm|1,0\rangle)$.}
\label{pics}

\end{figure}

In this work, we introduce a chip-based system comprised of a finite-size nanowire PC waveguide with a pair of embedded QDs at opposite ends. Figure~\ref{pics} shows a schematic of our proposal,  which can be fabricated  using current growth techniques, as well as an energy level diagram for a pair of embedded QDs.  The geometry exploits the large spontaneous emission (SE) enhancements and a near lossless waveguide mode of the NW PCs~\cite{Angelatos2014} to mediate inter-QD interactions.  Using a quantum master equation formalism centered on the  photonic Green function, we demonstrate that this system  can strongly couple a pair of qubits.    We also study the fluorescence spectrum emitted from the device, which displays signatures of nonlinear coupling via photon transport.  In particular, we introduce a unique regime of quantum electrodynamics (QED), where significant exchange splitting between QDs occurs and a ``Mollow nonuplet'' (i.e., with nine spectral peaks) is obtained.

\section{Theory}
\label{sec:theory}
  The coupling dynamics of $N$ QDs (treated as two-level  atoms in the dipole approximation) in an arbitrary medium with permittivity $\epsilon({\bf r},\omega)$ is governed by the  Hamiltonian~\cite{Dung2002}:
$\hat{H}=\int  d^3\mathbf{r} \int_0^\infty d\omega\hbar \omega \hat{\mathbf{f}}^\dagger (\mathbf{r}; \omega)\hat{\mathbf{f}}(\mathbf{r}; \omega) + 
\sum_{n} \hbar\omega_n \hat{\sigma}^+_{n}\hat{\sigma}^-_{n} 
 -\sum_{n}\!\int_0^\infty d\omega\big( \hat{\mathbf{d}}_n\cdot
 \hat{\mathbf{E}}(\mathbf{r}_n; \omega) + {\rm H.c.}\big),
$
where the $n$th QD is at position $\mathbf{r}_n$, with resonance $ \omega_n$, and the dipole operator $ \hat{\mathbf{d}}_n= \mathbf{d}_n (\hat{\sigma}^-_{n}+\hat{\sigma}^+_{n})$, with $\mathbf{d}_n$ the  dipole moment of QD $n$.  Here $\hat{\mathbf{f}}$ is  a vectorial bosonic field annihilation operator, related to the electric field operator via
$\hat{\mathbf{E}}(\mathbf{r}; \omega)=
 i \sqrt{\frac{\hbar}{\pi\epsilon_0}}\int d^3\mathbf{r}'\sqrt{\text{Im}\{\epsilon(\mathbf{r}'; \omega) \}}\mathbf{G}(\mathbf{r}, \mathbf{r}'; \omega)\cdot\hat{\mathbf{f}}(\mathbf{r}'; \omega)$. $\mathbf{G}(\mathbf{r}, \mathbf{r}'; \omega)$ is the electric field Green function, describing the system response at $\mathbf{r}$ to a point source at $\mathbf{r'}$: $\left [\nabla \times \nabla \times - \frac{\omega^2}{c^2} \epsilon(\mathbf{r}) \right ]\mathbf{G}(\mathbf{r},\mathbf{r}';\omega) =\frac{\omega^2}{c^2}{\mathbf{1}} \delta(\mathbf{r}-\mathbf{r'})$. 
This approach naturally handles lossy and open structures, and in the limit of $\text{Im}\{\epsilon(\mathbf{r}'; \omega)\}=0$, the properties of the Green tensor allows one to recover $\hat{\mathbf{E}}$ as a sum over field modes~\cite{Gruner1996}  We direct the reader to appendices \ref{subsec:G} and \ref{subsec:fieldquant} for a more thorough discussion of $\mathbf{G}$ and its relation to $\hat{\mathbf{E}}$.  Working in a rotating frame with respect to a laser frequency $\omega_L$, 
 we derive the quantum  master equation for the system of QDs interacting with a photonic reservoir, shown in Ref.~\onlinecite{Dung2002} and derived in detail in appendix \ref{subsec:me}. In the weak-coupling regime, with the system-reservoir coupling given by the dipole interaction in the rotating-wave approximation, we  apply  the second-order Born and Markov approximations to the interaction Hamiltonian, trace out the electromagnetic degrees of freedom, and after some algebra arrive at the master equation for the reduced density operator~\cite{Dung2002}:
\begin{align} 
\dot{{\rho}}=-&i\sum_n{\Delta\omega}_n[  \hat{\sigma}^+_{n}\hat{\sigma}^-_{n}, {\rho}]
-i\sum_{n, n'}^{n \neq n'} \delta_{n, n'}[ \hat{\sigma}^+_{n}\hat{\sigma}^-_{n'}, {\rho}] \nonumber \\ 
+&\sum_{n, n'}\Gamma_{n, n'}\left (\hat{\sigma}^-_{n'}{\rho}\hat{\sigma}^+_{n}-\frac{1}{2}\{\hat{\sigma}^+_{n}\hat{\sigma}^-_{n'}, {\rho}\}\right)  \nonumber \\ 
- &\frac{i}{\hbar}[\hat{H}_{\rm drive}, \hat{\rho}]+\sum_{n}\gamma'_{n}\mathcal{L}[\hat{\sigma}^+_{n}\hat{\sigma}^-_{n}],
\label{eq:LME}
\end{align} where $\Delta{\omega}_n =(\omega_n'-\omega_L)$, 
 $\omega_n' = \omega_n +\Delta_{n}$, and  
$\Delta_{n}=\frac{-1}{\hbar\epsilon_0}\mathbf{d}_n \cdot\text{Re}\left\{ \mathbf{G} ( \mathbf{r}_n, \mathbf{r}_{n};\omega_{n})\right\}\cdot \mathbf{d}_{n}$
is the photonic Lamb shift;  the inter-QD  coupling terms are
$\delta_{n, n'}|_{n \neq n'}=\frac{-1}{\hbar\epsilon_0}\mathbf{d}_n \cdot\text{Re}\left\{ \mathbf{G} ( \mathbf{r}_n, \mathbf{r}_{n'};\omega'_{n'})\right\}\cdot \mathbf{d}_{n'} $
and $\Gamma_{n, n'}= \frac{2}{\hbar\epsilon_0}\mathbf{d}_n\cdot \text{Im}\left\{ \mathbf{G} ( \mathbf{r}_n, \mathbf{r}_{n'}; \omega'_{n'})\right\}\cdot \mathbf{d}_{n'}.$  The pump term ${H}_{\rm drive} =\sum_n\frac{\hbar\Omega_{R,n}}{2}(\hat{\sigma}^+_{n}+ \hat{\sigma}^-_{n})$ represents the external coherent  drive  applied to each QD at laser frequency $\omega_L$,  where the effective Rabi field  $\Omega_{R,n}=\langle\hat{\mathbf{E}}_{{\rm pump}, n}(\mathbf{r}_n) \cdot\mathbf{d}_n\rangle/\hbar$~\cite{Carmichael1999}. 
In the above derivation, the Rabi fields and coupling terms (in units of frequency) are smaller than the frequency scale over which an appreciable change in the LDOS occurs, so that the scattering rates are essentially pump independent~\cite{Ge2013} and the Born and Markov approximations are  valid~\cite{Carmichael1999}.  For the PC system in this paper, the coupling rates are indeed well within the weak-coupling regime and Rabi fields were chosen to be of similar strength.  We  use the scattered part of the Green function and thus subtract off the divergent homogeneous vacuum Lamb shift, which is already included in $\omega_x$.  We note that for a nonzero pump, the rate terms $\delta_{n, n'}$ and $\Gamma_{n, n'}$  are evaluated at $\omega_L$ instead of $\omega'_{n'}$, although the LDOS is essentially flat over this frequency range.
To better highlight the radiative coupling dynamics, 
we also neglect pump-induced dephasing effects (e.g., through phonon-induced interactions).  However, the final term in Eq.~(\ref{eq:LME}) accounts for pure dephasing via the standard Lindbladian superoperator $\mathcal{L}[\hat{O}]= (\hat{O}{\rho}\hat{O}^\dagger-\frac{1}{2}\{\hat{O}^\dagger\hat{O}, {\rho}\})$, with $\gamma'_n$ the pure dephasing rate of QD $n$.
Importantly, Eq.~\eqref{eq:LME}  allows one to analyze the radiative coupling dynamics of a system of QDs in an arbitrary dielectric bath medium such as a PC waveguide, where all of the coupling depends explicitly on  the medium Green functions.

\section{Results}

\subsection{Proposed structure} 

\begin{figure*}[ht!]
\centering
\vspace{-5pt}
\includegraphics[width=\textwidth]{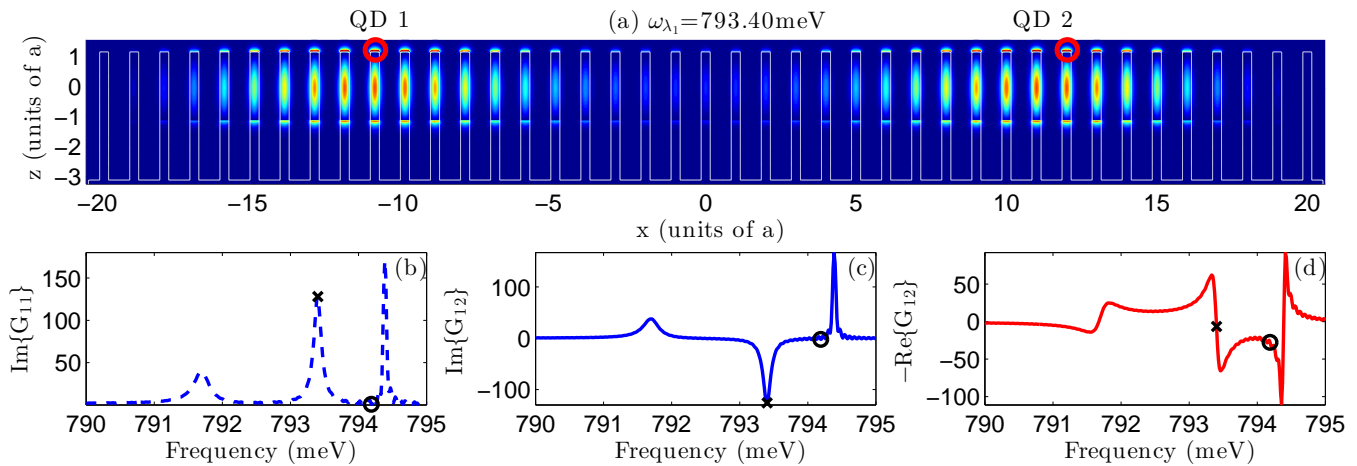}
\vspace{-20pt}
\caption{(Color online) (a)  $|\mathbf{E}_{\lambda_1}|^2$ the in $y\!=\!0$ plane of NW waveguide. QD locations are indicated via red circles.  (b)  $\text{Im}\{{G}(\mathbf{r}_{1/2}, \mathbf{r}_{1/2}; \omega)\}$, (c) $\text{Im}\{{G}(\mathbf{r}_{1}, \mathbf{r}_{2}; \omega)\}$, and (d) $-\text{Re}\{{G}(\mathbf{r}_{1}, \mathbf{r}_{2}; \omega)\}$, directly proportional to $\Gamma_{1,1}$, $\Gamma_{1,2}$, and $\delta_{1,2}$.  All rates are in units of $\text{Im}\{{G}^{\rm h}(\mathbf{r}, \mathbf{r}; {\omega})\}$.  The crosses  and circles indicate values at $\lambda_1$ and $\lambda_2$ respectively.}
\label{wgs}
\vspace{-10pt}
\end{figure*}

 Figure~\ref{pics}\subref{NWWG} shows a schematic of the proposed PC NW waveguide.  In our specific design, the waveguide has a length and width of 41$a$ and 7$a$, with lattice constant $a=0.5526\,\mu$m to produce a single vertically-polarized waveguide band with a mode edge near the  telecom wavelength of $1.550\,\mu$m.   As described previously~\cite{Angelatos2014}, a  waveguide is formed by reducing the radius of a single row of NWs, from $r_b=0.180a$ to $r_d=0.140a$. Light remains confined to the higher index  (GaAs, $\epsilon=13$) upper portion (height 2.27$a$) of the NWs, and the lower index portion (AlO, $\epsilon=3.1$, height 2$a$) separates the NWs from the substrate.  We consider a pair of vertically-polarized QDs that are embedded post-process at the top of selected NWs, where they efficiently couple into the waveguide Bloch mode antinode as depicted in the inset of Fig.~\ref{pics}\subref{NWWG}.  Each QD  resides on top of a NW ten unit cells from the center of the structure (separated by 21$a$, 10.6$\,\mu {\rm m}$). The relevant $\mathbf{G}$ components ($G=\mathbf{e}_z\cdot\mathbf{G}\cdot \mathbf{e}_z$) for the two QDs indicated, found using  a  finite-difference time-domain (FDTD) approach~\cite{Yao2009, LumericalSolutions}, are shown in Fig.~\ref{wgs}(b-d) through the waveguide band.  In a homogeneous medium with refractive index $n^{\rm h}_d$,  one can derive the Green function analytically, and  the spontaneous emission rate will be directly proportional to $\text{Im}\{{G}^{\rm h}(\mathbf{r}, \mathbf{r}; {\omega})\}=\omega^3 n^{\rm h}_d/(6\pi c^3)$~\cite{Novotny2006}.  All $\mathbf{G}$ components are thus given in units of $ \text{Im}\{{G}^{\rm h}(\mathbf{r}, \mathbf{r}; {\omega})\}$ to highlight the  rate enhancements present in this system relative to a pair of QDs in free space.  We note that the largest LDOS peak corresponds to the quasimode formed at the mode edge of a slow-light waveguide mode~\cite{Angelatos2014}, whereas the lower frequency peaks are Fabry\textendash{}P\'erot ripples due to facet reflections~\cite{Angelatos2014}. Optimal coupling is achieved by choosing the mode which maximizes the symmetric photon exchange terms, $|\text{Im}\{\mathbf{G}(\mathbf{r}_{1/2}, \mathbf{r}_{2/1}; \omega)\}|$.  The photonic mode $ \lambda_1$ which best achieves this is shown in Fig.~\ref{wgs}(a),  containing antinodes at the symmetric QD positions.

\subsection{Free evolution case}
  We first study the dynamics of a single excited QD (QD 1) with no external drive.  Both QDs were taken to have a vertical dipole moment of ${d}=30\,$D ($0.626\,
 $e-nm) and a renormalized exciton line at ${\omega}_x'=\omega_{\lambda_1}=793.40$\,meV.  We also include a pure dephasing rate of $1\,\mu$eV in all calculations, similar to experimental numbers on InAs QDs at 4\,K~\cite{Weiler2012}.  The calculated SE rates and exchange terms (in units of $\text{Im}\{{G}^{\rm h}(\mathbf{r}, \mathbf{r}; {\omega}_x')\}$) are 131.7 and 129.8, respectively, for the chosen positions and frequencies ($22.40$ and $22.05\,\mu$eV).  This large coupling rate is  remarkable given the  openness of the structure and the large spatial separation of the QDs, and exceeds that found in comparable slab PC waveguides \cite{Minkov2013}.  

\begin{figure}[t]
\captionsetup[subfigure]{labelformat=empty}
\vspace{-15pt}
\subfloat[\vspace{-10pt}]{\label{qdyn1}\includegraphics[width=0.24\textwidth]{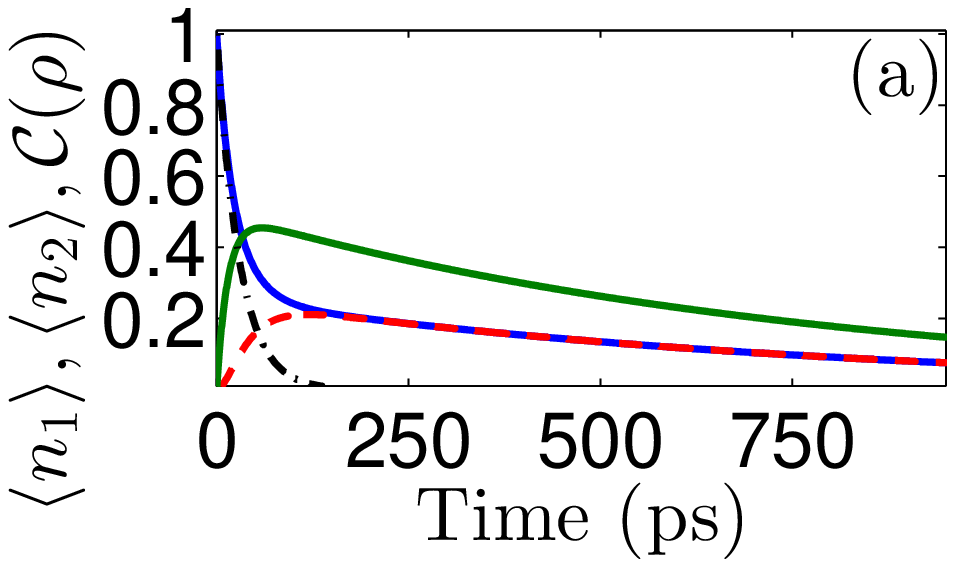}}
\subfloat[\vspace{-10pt}]{\label{qdynsym}\includegraphics[width=0.24\textwidth]{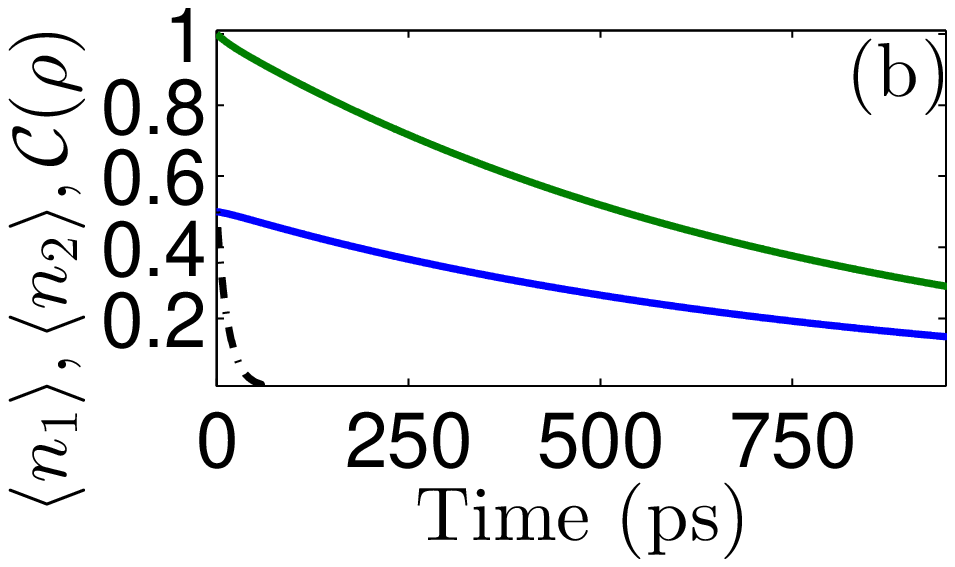}}
\vspace{-5pt}
\caption{(Color online) (a) Coupling dynamics between two QDs, with QD 1 initially excited.  Population of QD 1 (2) is shown in dark blue (dashed red) and entanglement in light green.  The dash-dotted black line displays the population of QD 1 in the same structure without QD 2, for comparison. (b) Dynamics of initially entangled pair state.  Single QD population of state initialized in $|\psi_+\rangle$ ($|\psi_-\rangle$) in dark blue (dash-dotted black) and concurrence of  $|\psi_+\rangle$ in light green.}
\label{dyn}
\end{figure}

Having calculated the relevant photonic Green functions, we solve the master equation (Eq.~\eqref{eq:LME}) for the density matrix $\rho(t)$~\cite{Tan1999}, which is used to obtain the population of each QD from $\langle\hat{n}_n(t)\rangle=\text{Tr}\{ \hat{\sigma}^+_n\hat{\sigma}^-_n \rho\}$.  To measure the entanglement between the pair of QDs, we calculate the system concurrence $\mathcal{C}(\rho)= \text{max}\{0, \lambda_1-\lambda_2-\lambda_3-\lambda_4\}$~\cite{Wootters1998}.  $\lambda_i$ are the eigenvalues of $\sqrt{\sqrt{\rho}\tilde{\rho}\sqrt{\rho}}$ in decreasing order, where the spin-flipped density matrix is defined as $\tilde{\rho}=\hat{\sigma}_{y, 2}\hat{\sigma}_{y, 1}\rho^*\hat{\sigma}_{y, 1}\hat{\sigma}_{y, 2}$.  The concurrence ranges from zero  for a separable state up to one for an ideal Bell state and increases monotonically with entanglement of formation; a state with non-negligible concurrence is considered entangled~\cite{Wootters1998}. As a consequence of the  weak-coupling regime, $\mathcal{C}(\rho)\leq0.5$ for a pair of identical QDs with one initially excited~\cite{Dung2002a}. The populations $\langle{n}_n\rangle$  and $\mathcal{C}(\rho)$ for QD 1 initially excited  are shown in Fig.~\ref{dyn}\subref{qdyn1}, with a long-lived entangled state clearly forming as the QDs couple resonantly to the waveguide mode and exchange their single excitation.  The system is seen to  remain populated far longer than in the single QD case; the lifetime of the entanglement also exceeds that of a comparable QD-PC system in the strong-coupling regime, where the entanglement falls to zero after $200$\,ps~\cite{Yao2009bell}.  When compared to  an idealized QD-plasmon waveguide system~\cite{Gonzalez-Tudela2011}, we achieve a higher peak entanglement and similar lifetime.

We next examine the dynamics of the symmetric and asymmetric entangled states $|\psi_{\pm}\rangle=\frac{1}{\sqrt{2}}(|0,1\rangle\pm|1,0\rangle)$, where the first (second) quantum number refers to the first (second) QD.  For this system, $|\psi_{\pm}\rangle$ populations decay at $\Gamma_\pm=\Gamma_{1,1}\pm\Gamma_{1,2}$~\cite{Dung2002, Yao2009bell}.  Due to the phase difference in the effective Bloch mode between the two QD positions, $\Gamma_{12}$ is negative,  extending the lifetime of $|\psi_+\rangle$ ($1/\Gamma_+\approx 50/\Gamma_-$) as shown in Fig.~\ref{dyn}\subref{qdynsym}.  We note that this lifetime exceeds that found for a symmetric state in the idealized plasmon waveguide structure~\cite{Gonzalez-Tudela2011}, and that by changing the positions or resonances of the QDs one can invert this relationship such that the asymmetric state will have the drastically longer lifetime, a feature that has useful quantum information applications.

\subsection{Field driven case} 
 To investigate nonlinear coupling we  consider a pump field applied to QD 1 via $\hat{H}_{\rm drive}$, and calculate the resulting spectrum by taking the Fourier transform of the correlation function.  This approach maintains the fermionic nature of the QDs, fully including saturation and nonlinear effects~\cite{Carmichael1999} and is derived in appendix \ref{subsec:spec}.  The total incoherent spectrum measured by a point detector at position $\mathbf{r}_D$, incorporating filtering via light propagation, is $S_{D}(\omega)=S_{D,1}+S_{D,2}+S_{D}^{\rm int}$, where $S_{D, n}= |\mathbf{G}(\mathbf{r}_D, \mathbf{r}_n; \omega)\cdot\mathbf{d}_n/\epsilon_0|^2 {\rm Re}\{ S^0_{ n, n}(\omega)\}$~\cite{Ge2013} and $S^0_{n,n'}(\omega)$ is related to the traditional incoherent spectrum from a two-level atom~\cite{Carmichael1999}: $S^0_{n, n'}(\omega)=\lim_{t \to \infty} \int_0^\infty d\tau ( \langle\hat{\sigma}^+_n (t+\tau) \hat{\sigma}^-_{n'}(t)\rangle-\langle\hat{ \sigma }^+_n(t)\rangle\langle\hat{\sigma}^-_{n'}(t)\rangle ) e^{ i(\omega_L - \omega ) \tau }$. The interference term $S_{D}^{\rm int}={\rm Re}\{g_{1,2}S^0_{1,2}+g^*_{1,2}S^0_{2,1}\}$, where the coupling term $g_{1,2}=\frac{1}{\epsilon_0^2}\mathbf{d}_1\cdot\mathbf{G}^*(\mathbf{r}_1, \mathbf{r}_D; \omega)\cdot\mathbf{G}(\mathbf{r}_D, \mathbf{r}_2; \omega)\cdot\mathbf{d}_2$ \cite{Note1}.  Due to  rapid exchange  between QDs in this system, $S_{D}^{\rm int}$ does not contain any interesting retardation-related interference effects.  The system behavior is best understood in terms of  individual QD spectra $S_{D,n}$, which can be isolated in measurements by choosing  $\mathbf{r}_D$ appropriately.

\begin{figure}[t]
\captionsetup[subfigure]{labelformat=empty}
\vspace{-15pt}
\subfloat[\vspace{-10pt}]{\label{qdyn}\includegraphics[width=0.24\textwidth]{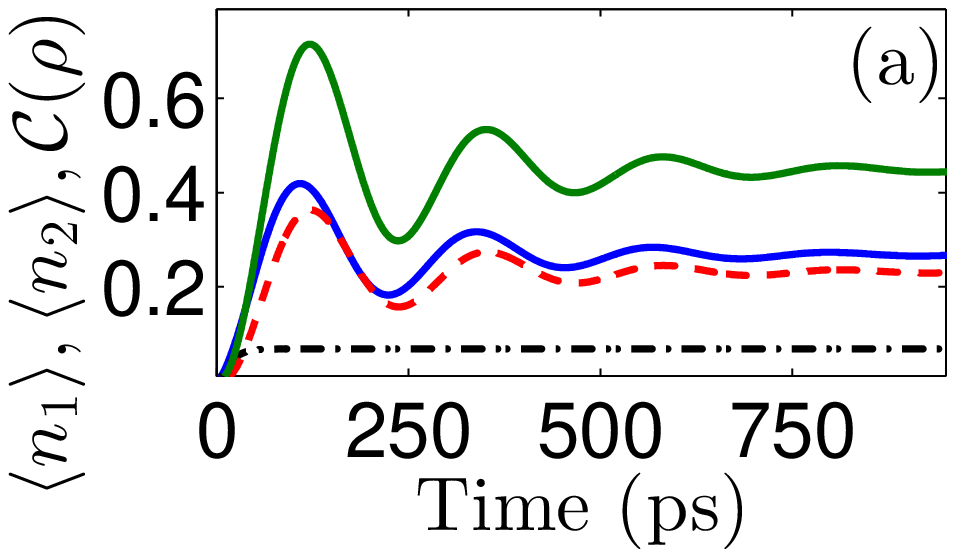}}
\subfloat[\vspace{-10pt}]{\label{spec}\includegraphics[width=0.24\textwidth]{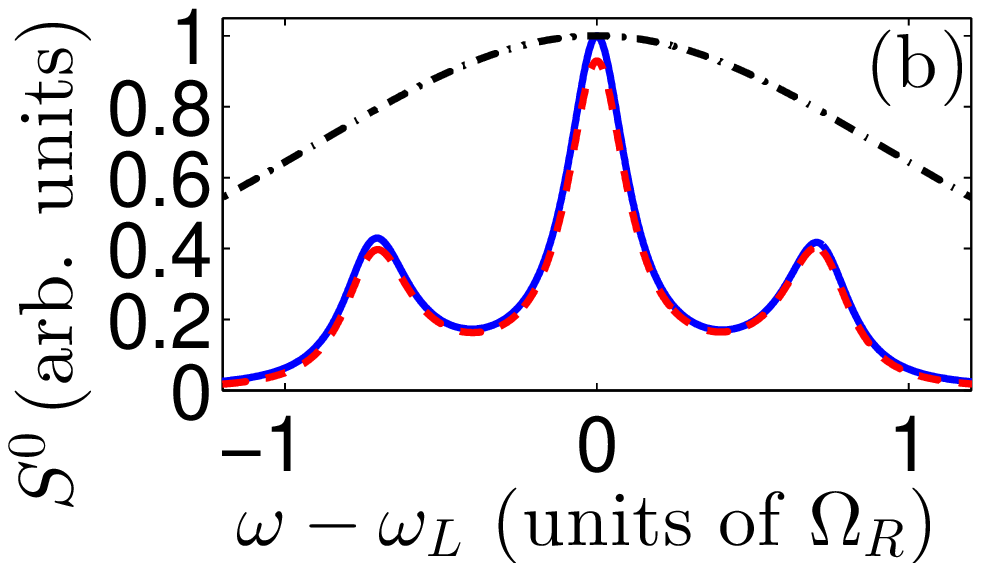}}
\vspace{-5pt}
\caption{(Color online) (a) Population and concurrence when QD 1 is driven, following the  labeling convention of Fig.~\ref{dyn}(a).  (b) Detected spectrum from QD 1 (2) in solid blue (dashed red).  The QD 1 spectrum  of an identical system without QD 2 is shown in dash-dotted black. A Mollow triplet is only observed for the two QD system under this excitation condition.}
\label{pdyn1}
\vspace{-15pt}
\end{figure}

 Figure~\ref{pdyn1}\subref{qdyn} shows the system dynamics with QD 1 driven by a $\Omega_R=25\,\mu$eV  pump at $\omega_L=\omega_{\lambda_1}+\delta_{1,2}$ (both QDs are initially in the ground state).  The dipole moment has been increased to an experimentally accessible 60\,Debye to better highlight exchange effects, but all other parameters remain the same.  It can be seen that a highly entangled state is formed with steady state $ \langle{n}_1\rangle$, $ \langle{n}_2\rangle$, and  $\mathcal{C}(\rho)$ of 0.27, 0.23, and 0.45, respectively.  We note that the strong medium-assisted photon exchange leads to Rabi oscillations and steady-state populations in the unpumped QD 2 almost identical to that of QD 1, and the chosen $\omega_L$ maximizes the steady-state $\mathcal{C}(\rho)$.  Figure~\ref{pdyn1}\subref{spec} displays the incoherent spectra of both QD 1 and QD 2, as well as that of an identical system containing only QD 1.  We  show $S^0$, but assume detector positions directly above each QD, where $|\mathbf{G}(\mathbf{r}_{D}, \mathbf{r}_n; \omega)|\gg|\mathbf{G}(\mathbf{r}_{D}, \mathbf{r}_{n'}; \omega)|$ such that emission from a single QD dominates and $S_D\propto S^0_{n,n}$.  The Mollow triplet, a clear signature of a driven fermionic system,  is observed in both QDs despite the lack of external Rabi field on QD 2.   The dynamics are dominated by the $\Gamma$ exchange terms, with $\Gamma_{1,2}=\Gamma_{2,1}=-88.2\,\mu$eV, and $\Omega_R=25\,\mu$eV.  In addition, the sideband splitting has been reduced from the traditional $\Omega_R$ due to this resonant photon exchange, with $\Delta_{R, 2}=0.701\Omega_R$ and $\Delta_{R, 1}=0.704\Omega_R$.  In particular, the Rabi field seen by QD 2 is due entirely to photons emitted from QD 1 via $\Gamma_{2,1}$, and the Rabi field at QD 1 is similarly dominated by the $\Gamma_{1,2}$ process, although it has been increased slightly by the pump.  As the position and intensity Mollow sidebands are directly dependent on the strength of the pump and exchange terms, one can experimentally study the coupling dynamics of this system by measuring the spectra emitted from each QD when the other is pumped.

 \subsection{Strong exchange regime} 
\begin{figure}[t]
\captionsetup[subfigure]{labelformat=empty}
\subfloat[\vspace{-10pt}]{\label{levels}\includegraphics[width=0.16\textwidth]{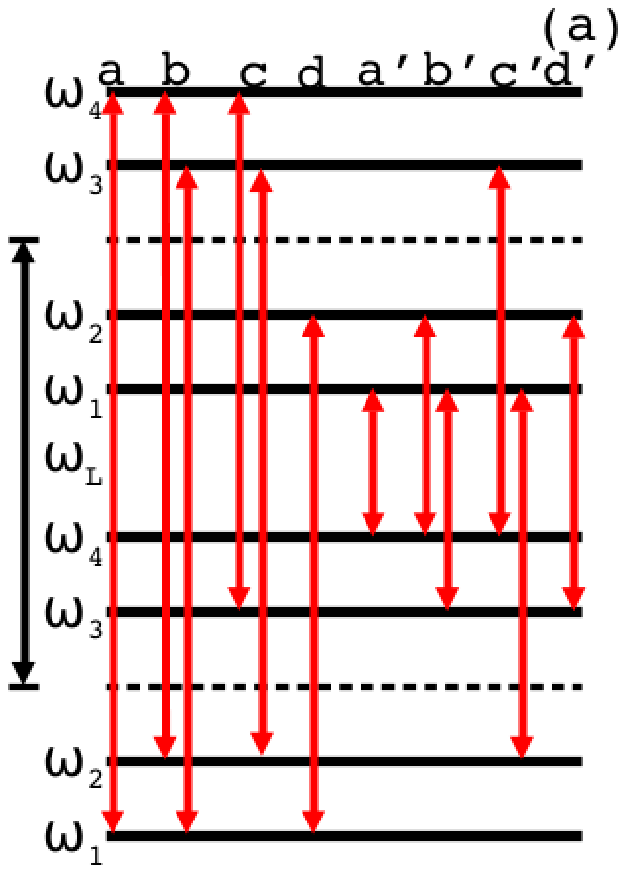}}
\subfloat[\vspace{-10pt}]{\label{peaks}\includegraphics[width=0.33\textwidth]{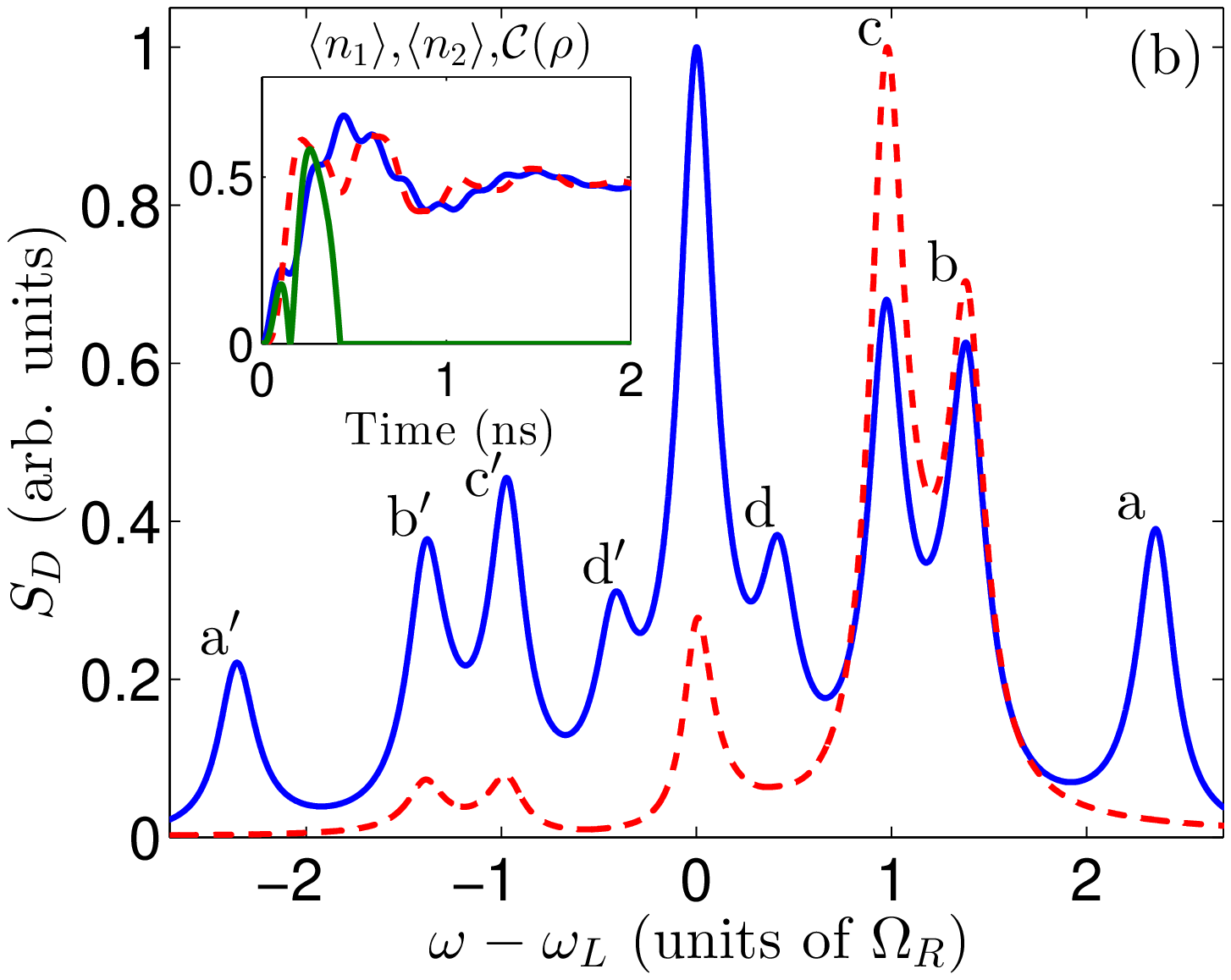}}
\vspace{-5pt}
\caption{(Color online) (a)  Energy levels and expected transitions for system evolving under ${H}_{\rm eff}$.  Unprimed transitions are from the interaction picture, and primed are found when one considers the full Hamiltonian.  The  transitions between identical levels at $\omega_L$ are not labeled.  (b)  The detected spectrum from QD 1 (2) in solid blue (dashed red).  The populations and concurrence are shown in the inset and again follow the convention of Fig.~\ref{dyn}(a). The $a$ and $d$ transitions do not appear in the QD 2 spectrum, as they have no effect on the state of QD 2.  $|\mathbf{G}(\mathbf{r}_D, \mathbf{r}_n; \omega)|$ increases with $\omega$ near $\omega_L$, causing spectral peak amplitudes to increase as well.}
\label{sys}
\end{figure}

 Lastly, we study a system where in contrast with the previous cases, we work in a regime with $\delta_{1,2}\gg\Gamma_{1,1}$, by choosing $\omega'_x=\omega_{\lambda_2}=794.19\,$meV and ${d}=60$\,D.  This allows dipole-dipole coupling, or exchange splitting, to control the system dynamics with $\delta_{2, 1}=\delta_{1,2}=-9.68\,\mu$eV, $\Gamma_{1,1}=0.64\,\mu$eV, $\Gamma_{1,2}=\Gamma_{2,1}=0.41\,\mu$eV. We note that this exchange splitting is on the order of that reported for neighboring QDs~\cite{Unold2005} or QDs in a shared cavity \cite{Laucht2010}, despite the large spatial separation in our device. In this regime, the system evolves under $\hat{H}_{\rm eff}= \hbar\Delta\omega_{\lambda_2} (\hat{\sigma}^+_{1}\hat{\sigma}^-_{1}+\hat{\sigma}^+_{2}\hat{\sigma}^-_{2}) +\hbar\delta_{1,2}\left(\hat{\sigma}^+_{1}\hat{\sigma}^-_{2}+\hat{\sigma}^+_{2}\hat{\sigma}^-_{1}\right)
+\hbar\frac{\Omega_R}{2}\left( \hat{\sigma}^+_{1}+ \hat{\sigma}^-_{1}\right)$.
With no pump the eigenstates of $\hat{H}_{\rm eff}$ are simply $|0,0\rangle$, $|\psi_\pm\rangle$, and $|1,1\rangle$ with energies 0, $\hbar\omega_{\lambda_2}\pm\hbar\delta_{1,2}$, and $2\hbar\omega_{\lambda_2}$ respectively, mimicking a biexcitonic cascade system with level splitting as shown in Fig.~\ref{Elev1}.  For non-zero pump, we find a Stark-shifted level structure, $E_i/\hbar=\Delta\omega_{\lambda_2}\pm \frac{1}{2}\sqrt{A\pm B }$, where $A=2\delta_{1,2}^2 +2\Delta\omega_{\lambda_2}^2+\Omega_R^2$, and $B= 2\sqrt{ \delta_{1,2}^4+\delta_{1,2}^2 \Omega_R^2-2\delta_{1,2}^2 \Delta\omega_{\lambda_2}^2 +\Omega_R^2\Delta\omega_{\lambda_2}^2+ \Delta\omega_{\lambda_2}^4 }$.   The temporal periodicity of the original Hamiltonian allows one to treat it in the Floquet picture, resulting in an infinite sequence of the interaction picture energy levels centered at $n\hbar\omega_L$, where $n$ is an integer.  Since only QD 1 is driven, we can truncate this sequence to the $n=0$ and $n=1$ sets, corresponding to the absorption of 0 or 1 photons from the laser.  The resultant energy levels are shown in Fig.~\ref{sys}\subref{levels}, where the four unique transitions of the interaction picture are labeled $a$-$d$.  This leads to a nine-peaked observable spectrum of the full time-dependent Hamiltonian, with the ninth peak being a four-fold degenerate transition at $\omega_L$.

We solved the dynamics of the above system with $\Omega_R=10\,\mu$eV and $\Delta\omega_{\lambda_2}=0$, and the resulting detectable spectrum, populations, and concurrence are shown in Fig.~\ref{sys}\subref{peaks}.  The  dressed energy levels are calculated to be  $E_i=\pm1.18\Omega_R,\,\pm0.212\Omega_R$, indexed by increasing energy, and clear signatures of all the expected transitions are observed. $|\psi_1\rangle$ and $|\psi_4\rangle$ are anticorrelated exciton states which behave similar to (and converge to) $|\psi_+\rangle$ and $|\psi_-\rangle$ respectively, whereas $|\psi_2\rangle$ and $|\psi_3\rangle$ are asymmetric and symmetric combinations of the biexciton and vacuum state, yielding for zero pump correlated exciton states $\frac{1 }{\sqrt{2}}(|1,1\rangle\mp|0,0\rangle)$.  These results are robust with respect to pure dephasing; we stress that we are using an experimentally viable $\gamma'=1\,\mu$eV in the above work, and a numerical study has indicated that these peaks remain resolvable up to $\gamma'\simeq5\mu$eV.

Of particular importance for this system is the  spatial filtering of respective QDs.  Specifically, we define a pair of detectors $D$ and $D'$ placed at mirror positions 1$a$ ($0.5526\,\mu {\rm  m}$) from the terminus of structure, along $y=z=0$ (see Fig.~\ref{wgs}(a)) with $D$ closer to QD 2 and $D'$ closer to QD 1.  Throughout the frequency range of interest, $|\mathbf{G}(\mathbf{r}_D, \mathbf{r}_1; \omega)|=|\mathbf{G}(\mathbf{r}_{D'}, \mathbf{r}_2; \omega)|\approx 8|\mathbf{G}(\mathbf{r}_{D'}, \mathbf{r}_1; \omega)|=8|\mathbf{G}(\mathbf{r}_D, \mathbf{r}_2; \omega)|$.  In consequence, the spectra of QDs 1 and 2 in Fig.~\ref{sys}\subref{peaks} correspond almost exactly with the total spectra observed at $D$ and $D'$ respectively, indicating that QDs can be studied individually by taking advantage of the inherent  structural  filtering.  We stress that one cannot isolate individual QD spectra in a comparable cavity structure, with the strong spatial filtering originating from the rich LDOS of the finite-sized waveguide.  Furthermore, this exchange-splitting regime is wholly inaccessible  in a simple cavity structure, as $\text{Re}\{\mathbf{G}\}$ falls off rapidly away from the peak of a Lorentzian LDOS, resulting in dynamics which are unavoidably dominated by the more rapid $\Gamma$ processes.    As such, the multiple-peak spectra of Fig.~\ref{sys}\subref{peaks} is inherent to our proposed PC waveguide structure.  This ability to model an effective four-level system and separately observe  each component indicates that these structures could potentially serve as many-body simulators or to study multi-QD quantum dynamics. Indeed, this type of device could readily be scaled up to $n$ QDs to simulate a $2^n$-level system. Furthermore, the ability to achieve substantial exchange splitting without relying on a strongly-coupled cavity is quite remarkable and could possibly be exploited to produce a CNOT gate.

\section{Conclusion} 
We have analyzed and explored the quantum dynamics of a pair of QDs embedded in a NW PC, including realistic factors such as finite-sized effects and radiative loss. We found that this system can access a broad range of quantum dynamical regimes.  By maximizing radiative coupling, we demonstrated the formation of a highly entangled state and photon-exchange-dependent Mollow triplet in the spectrum of an unpumped QD.  We then showed that, through tuning the operating frequency, one can control the effective system Hamiltonian and simulate a variety of quantum systems.  In particular, we discussed a unique quantum optical regime which produces nine signature spectral peaks.  This versatility makes these structures attractive for use in quantum information science and to explore quantum optics on chip. 

 \section{Acknowledgements}  
This work was supported by the Natural Sciences and Engineering Research Council of Canada and Queen's University.

\renewcommand{\thefigure}{A\arabic{figure}}  
\renewcommand{\thesubsection}{A\arabic{subsection}}   
\renewcommand{\theequation}{A\arabic{equation}}  
\setcounter{figure}{0}    
\setcounter{section}{0}    
\setcounter{equation}{0}    
\section*{Appendix}
\subsection{Classical Electric field Green functions}
\label{subsec:G}
In an arbitrary inhomogeneous linear medium, the classical electric field  obeys the partial differential equation~\cite{Novotny2006}
\be
\left[\nabla \times \mu^{-1}(\mathbf{r}; \omega) \nabla \times - \frac{\omega^2}{c^2}\epsilon(\mathbf{r}; \omega)\right]\mathbf{ E}(\mathbf{r}; \omega)=i\omega\mu_0\mathbf{j}_s(\mathbf{r}; \omega),
\ee
where $c$ is the speed of light, $\mu$ and $\epsilon$ are the material permeability and permittivity, 
%($\epsilon$ including the conduction current), 
and $\mathbf{j}_s$ is the noise current source.  We have also assumed isotropy, although anisotropy can be accounted for by simply substituting the scalar material parameters with their tensorial counterparts: $\epsilon, \mu \to \boldsymbol\epsilon, \boldsymbol\mu$.  In this work, we are concerned with non-magnetic materials and polarization sources, so we take $\mu=1$ and $\mathbf{j}_s=-i\omega\mathbf{P}_s$, yielding
\be
\left[\nabla \times\nabla \times - \frac{\omega^2}{c^2}\epsilon(\mathbf{r}; \omega)\right]\mathbf{E}(\mathbf{r}; \omega)= \frac{\omega^2}{c^2}\frac{\mathbf{P}_s(\mathbf{r}; \omega)}{\epsilon_0}.
\label{eq:helm}
\ee
The noise polarization source is the polarization associated with material absorption, included via the constitutive relation for electric displacement: $\mathbf{D}(\mathbf{r}; \omega)=\epsilon_0\epsilon(\mathbf{r})\mathbf{E}(\mathbf{r}; \omega)+\mathbf{P}_s(\mathbf{r}; \omega)$ \cite{Knoll2000}.  This is equivalent to perturbing the system permittivity $\epsilon\rightarrow\epsilon+\Delta\epsilon$, with $\mathbf{P}_s(\mathbf{r}; \omega)= \Delta\epsilon(\mathbf{r})\mathbf{E}(\mathbf{r}; \omega)$.  The inhomogeneous Helmoltz equation above is most readily solved through the Green function approach.  Specifically, we define the electric field Green tensor as the solution to
\be
\left [\nabla \times \nabla \times - \frac{\omega^2}{c^2} \epsilon(\mathbf{r})\\ \right ]\mathbf{G}(\mathbf{r},\mathbf{r}';\omega) =\frac{\omega^2}{c^2}{\mathbf{1}} \delta(\mathbf{r}-\mathbf{r'}),
\label{eg:Gdef}
\ee
where $\mathbf{1}$ is the unit dyad.  $\mathbf{G}$ is the electric field response at $\mathbf{r}$ to a point source at $\mathbf{r}'$.  We note that our $\mathbf{G}$ includes an additional factor of $\frac{\omega^2}{c^2}$ relative to other common sources \cite{Novotny2006, Knoll2000} in order to simplify a number of subsequent relations, this choice yields a $\mathbf{G}$ with units of inverse volume and is consistent with the Green function from a dipole source in the full Maxwell curl equations, which is more suited for numerical calculations or an arbitrary structure.   Once $\mathbf{G}$ is determined, by postmultiplying Eq.~\eqref{eg:Gdef} by $\frac{\mathbf{P}_s(\mathbf{r}; \omega)}{\epsilon_0}$ and integrating we find
\be
\mathbf{E}(\mathbf{r}; \omega)=\int_V d\mathbf{r}'\mathbf{G}(\mathbf{r}, \mathbf{r}'; \omega)\cdot\frac{\mathbf{P}_s(\mathbf{r}'; \omega)}{\epsilon_0},
\label{eq:EfromG}
\ee
where the integral is over the source volume.  
 We can always add the homogeneous solution $\mathbf{E}_0(\mathbf{r}; \omega)$, which satisfies $\nabla \times \nabla \times\mathbf{E}_0=\frac{\omega^2}{c^2} \epsilon(\mathbf{r})\mathbf{E}_0$, to the particular solution of   Eq.~\eqref{eq:EfromG}  so that the general solution is 
\be
\mathbf{E}(\mathbf{r}; \omega)=\mathbf{E}_0(\mathbf{r}; \omega)+\int_V d\mathbf{r}'\mathbf{G}(\mathbf{r}, \mathbf{r}'; \omega)\cdot\frac{\mathbf{P}_s(\mathbf{r}'; \omega)}{\epsilon_0}.
\label{eq:EfromG2}
\ee
The Green tensor has the following useful properties, all of which are proven in Ref.~\onlinecite{Knoll2000},
\begin{gather}
G_{j,i}(\mathbf{r}', \mathbf{r}; \omega)=G_{i,j}(\mathbf{r}, \mathbf{r}'; \omega),\label{eq:prop1}\\
\mathbf{G}^*(\mathbf{r}, \mathbf{r}'; \omega)=\mathbf{G}(\mathbf{r}, \mathbf{r}'; -\omega),\\
\int d\mathbf{r}''\epsilon_I(\mathbf{r}''; \omega)\mathbf{G}(\mathbf{r}, \mathbf{r}''; \omega)\cdot\mathbf{G}^*(\mathbf{r}'', \mathbf{r}'; \omega)=\text{Im}\{\mathbf{G}(\mathbf{r}, \mathbf{r}'; \omega)\}, \label{eq:prop3}
\end{gather}
where subscripts correspond to directional indices and we use the notation $\epsilon_I=\text{Im}\{\epsilon\}$ in what follows.

\subsection{Field Quantization}
\label{subsec:fieldquant}
We now proceed to quantize the electromagnetic field, considering a quantum electric field operator which is governed by the same Maxwell equations as its classical counterpart, and thus following Eq.~\eqref{eq:helm}.  Following the standard canonical quantization procedure, the fundamental system variables $\hat{\mathbf{f}}(\mathbf{r}; \omega)$~\cite{Knoll2000} become a continuous set of bosonic field annihilation operators which obey commutation relations $[\hat{{f}}_{j}(\mathbf{r}; \omega), \hat{{f}}_{j'}^\dagger(\mathbf{r}'; \omega')]=\delta_{j, j'}\delta(\mathbf{r}-\mathbf{r'})\delta(\omega-\omega')$ and $[\hat{{f}}_{j}(\mathbf{r}; \omega), \hat{{f}}_{j'}(\mathbf{r}'; \omega')]=0$~\cite{Suttorp2004}.  We note that this is done in the Schr\"{o}dinger picture and $\omega$ indicates that  $\hat{\mathbf{f}}(\mathbf{r}; \omega)$ is associated with the mode $\omega$ and not the Fourier transform of the time variable $t$.  It can be shown that the noise polarization excites these modes through~\cite{Knoll2000, Suttorp2004}
\be
\hat{\mathbf{P}}_s(\mathbf{r}; \omega)=-i \sqrt{\frac{\hbar \epsilon_0\epsilon_I(\mathbf{r}; \omega)}{\pi}}\hat{\mathbf{f}}(\mathbf{r}; \omega).
\label{eq:Psq}
\ee
$\hat{\mathbf{P}}_s$ generates the quantized electric field operator via a quantum version of Eq.~\eqref{eq:EfromG}, and the electric field operator is thus given by
\be
\hat{\mathbf{E}}(\mathbf{r}; \omega)=
 i \sqrt{\frac{\hbar}{\pi\epsilon_0}}\int d\mathbf{r}'\sqrt{\epsilon_I(\mathbf{r}'; \omega) }\mathbf{G}(\mathbf{r}, \mathbf{r}'; \omega)\cdot\hat{\mathbf{f}}(\mathbf{r}'; \omega),
\label{eq:Eq}
\ee
where the integral is over all space and the homogeneous contribution is included in $\mathbf{f}$, as will become apparent in Sec.~\ref{sec:spec}.  Somewhat remarkably, $\mathbf{G}$ is the same Green function found classically via Eq.~\eqref{eg:Gdef}, and although this result could be deduced phenomenologically via Eqs.~\eqref{eq:EfromG} and \eqref{eq:Psq}, it is in fact rigorously justified~\cite{Suttorp2004}.  The total electric field operator is found via integration over $\omega$
\be
\hat{\mathbf{E}}(\mathbf{r})=\int_0^\infty d\omega\hat{\mathbf{E}}(\mathbf{r}; \omega)+\text{H.c.}=\hat{\mathbf{E}}^+(\mathbf{r})+\hat{\mathbf{E}}^-(\mathbf{r}).
\ee
\subsection{Derivation of the master equation}
\label{subsec:me}
In this section, we present a  derivation of Eq.~\eqref{eq:LME} of the main text.  We  use the traditional open quantum systems approach, deriving a master equation for the reduced density matrix of the system by applying the standard Born and Markov approximations and tracing over the reservoir to produce a series of Lindbladian terms.  This route is taken with the hope that readers will find the process familiar and the approximations made will be more transparent.  An excellent alternative derivation is presented in Ref.~\onlinecite{Dung2002}, which one can quickly see gives the same result if the coherent pump is included in the system Hamiltonian;  the compatibility of these separate approaches further justifies our final result.

As presented in Sec.~\ref{sec:theory}, a system of two-level atoms interacting with the surrounding electromagnetic environment in the dipole approximation is governed by the Hamiltonian~\cite{Dung2002}:
\begin{align}
\hat{H}=&\int  d^3\mathbf{r} \int_0^\infty d\omega\,\hbar \omega \hat{\mathbf{f}}^\dagger (\mathbf{r}; \omega)\hat{\mathbf{f}}(\mathbf{r}; \omega) + 
\sum_{n} \hbar\omega_n \hat{\sigma}^+_{n}\hat{\sigma}^-_{n} \nonumber \\
 &-\sum_{n}\!\int_0^\infty d\omega\big( \hat{\mathbf{d}}_n
 \cdot\hat{\mathbf{E}}(\mathbf{r}_n; \omega) + {\rm H.c.}\big).
\label{eq:H}
\end{align}
 We begin by separating Eq.~\eqref{eq:H} into the emitter system, photonic reservoir, and interaction components, $\hat{H}=\hat{H}_S+\hat{H}_R+\hat{H}_{SR}$, and modify it to include the possibility of a continuous-wave pump applied to each emitter.  This coherent drive is included in the system Hamiltonian as $\hat{H}_{\rm drive}=\sum_n \frac{1}{2}\mathbf{E}_{{\rm pump}, n}(\mathbf{r}_n)\cdot\mathbf{d}_n(\hat{\sigma}^+_{n}e^{-i\omega_Lt}+ \hat{\sigma}^-_{n}e^{i\omega_Lt})$~\cite{Carmichael1999}, where due to its large amplitude we treat the drive field as a $c$-number and ignore fluctuations: $\mathbf{E}_{\rm pump}=\langle\hat{\mathbf{E}}_{\rm pump}\rangle$. The drive term  $\hat{H}_{\rm drive}$ is simply the laser field contribution to the dipole interaction term, and we define the effective Rabi field as $\Omega_{R,n}=\langle\hat{\mathbf{E}}_{{\rm pump}, n}(\mathbf{r}_n)\rangle\cdot\mathbf{d}_n/\hbar$.  We then transform to a frame rotating with laser frequency $\omega_L$ ($\hat{H}\to\hat{U}_L^\dagger(t) \hat{H} \hat{U}_L(t)$, $\hat{U}_L(t)=e^{-i\omega_L\sum_n\hat{\sigma}^+_{n}\hat{\sigma}^-_{n}t}$) and find system, reservoir, and interaction components of the Hamiltonian, defined through 
\begin{align}
\hat{H}_S=&\sum_{n} \hbar(\omega_n-\omega_L) \hat{\sigma}^+_{n}\hat{\sigma}^-_{n} +\hbar\frac{\Omega_{R,n}}{2}(\hat{\sigma}^+_{n}+ \hat{\sigma}^-_{n}) ,\\
{\hat{H}_R}=&\int  d^3\mathbf{r} \int_0^\infty d\omega\hbar \omega \hat{\mathbf{f}}^\dagger (\mathbf{r}; \omega)\hat{\mathbf{f}}(\mathbf{r}; \omega), \\
 \hat{H}_{SR}=&-\sum_{n}\left(\hat{\sigma}^+_{n}e^{i\omega_Lt}+ \hat{\sigma}^-_{n}e^{-i\omega_Lt}\right)\nonumber \\
~~&\times\int_0^\infty d\omega \big(\mathbf{d}_n\cdot
 \hat{\mathbf{E}}(\mathbf{r}_n; \omega) + {\rm H.c.}\big),
 \end{align}
 where we have expanded the dipole operator in the rotating frame.
 The density matrix of the total system and reservoir evolves according to the Schr\"{o}dinger equation $\dot{\rho}_T=\frac{1}{i\hbar}[\hat{H}, \rho_T]$.  We transform to the interaction picture (i.e., $\hat{O}_I=\hat{U}^\dagger(t)\hat{O}\hat{U}(t)$, $\hat{U}(t)=e^{-i(H_S+H_R)t/\hbar}$) where it is easily seen by combining the above two equations that the density matrix evolves as $\dot{\rho}_{T, I}=\frac{1}{i\hbar}[\hat{H}_{I}, \rho_{T,I}]$, with $H_I=H_{SR, I}$ for simplicity. We integrate to find 
 \be
 \rho_{T, I}(t)=\rho_I(0)R_{0}+\frac{1}{i\hbar}\int_0^t dt'[\hat{H}_{I}(t'), \rho_{T,I}(t')],
 \label{eq:rhoI}
 \ee
  where $R_{0}$ is the initial reservoir density matrix,  which we can always treat as a pure state \cite{Breuer2007}.  
 
 In the interaction picture, it is evident from the commutation relations discussed earlier that $\hat{\mathbf{f}} (\mathbf{r}; \omega, t)=\hat{\mathbf{f}} (\mathbf{r}; \omega)e^{-i\omega  t}$, whereas $\hat{\sigma}^{\pm}$ will be slowly varying, since $\Omega_R\ll\omega$ for optical frequencies and we are interested in resonant driving $\omega_L\approx\omega_n$.  We thus make the rotating-wave approximation in $H_{I}$, dropping the rapidly varying counter-rotating terms proportional to $\hat{\sigma}^+_{n}(t')\hat{\mathbf{f}}^\dagger(\mathbf{r}'; \omega)e^{i(\omega_L+\omega)t'}$ and  its Hermitian conjugate.  This is justified since the integration over $t'$ gives these terms a  factor of $\approx1/(\omega+\omega_L)$, and they are thus much smaller than the rotating-wave terms~\cite{Carmichael1999}.  To be explicit, we are using
 \be
  \hat{H}_{I}(t)=-\sum_{n}\int_0^\infty d\omega\hat{\sigma}^+_{n}(t) \mathbf{d}_n\cdot
 \hat{\mathbf{E}}(\mathbf{r}_n; \omega)e^{-i(\omega-\omega_L)t} %\nonumber \\
+ {\rm H.c.}
 \label{eq:rwa}
 \ee

 To produce an equation of motion for the system density matrix $\rho$ ($\rho=\text{Tr}_R\{\rho_T\}$), we insert $\rho_{T, I}(t)$ via Eq.~\eqref{eq:rhoI} into the interaction picture Schr\"{o}dinger equation and trace over the reservoir:
 \begin{align}
 \dot{\rho}_{I}=&\text{Tr}_R\{\frac{1}{i\hbar}[  \hat{H}_{I}, \rho_I(0)R_{0, I}]\}\nonumber \\
&-\frac{1}{\hbar^2}\int_0^t dt' \text{Tr}_R\{[  \hat{H}_{I}(t),[  \hat{H}_{I}(t'), \rho_{T,I}(t')]]\}.
\label{eq:ME1}
\end{align}
The above equation is simplified by a number of approximations.  We first take the mean initial system-reservoir coupling to be zero, such that $\text{Tr}_R\{\frac{1}{i\hbar}[  \hat{H}_{I}, \rho_I(0)R_{0}]\}=0$.  Even if this is not the case, the mean coupling with the system in  $R_0$ can simply be included as an additional term in the  Hamiltonian, such that the trace will indeed be zero in this renormalized system \cite{Carmichael1999}.  We then make the Born approximation, noting that the reservoir will be largely unaffected by its interaction with the system and assume the total density matrix evolves as: $\rho_T(t)=\rho(t)R_0+O(  \hat{H}_{SR})$~\cite{Carmichael1999}, and thus we do not need to iterate Eq.~\eqref{eq:ME1} into the Schr\"{o}dinger equation further.  Next, we  assume the evolution of the density matrix depends only on its current state and write Eq.~\eqref{eq:ME1} in time-convolutionless form (this is often referred to as the Born-Markov approximation)~\cite{Breuer2007, Ge2013},    
 \be
 \dot{\rho}_{I}=-\frac{1}{\hbar^2}\int_0^t d\tau \text{Tr}_R\{[  \hat{H}_{I}(t),[  \hat{H}_{I}(t-\tau), \rho_{I}R_0]]\},
\label{eq:ME2}
\ee
where $\rho_{I}=\rho_{I}(t)$.  The Born-Markov approximation is justified because the system dynamics are much slower than that of the bath; the system-reservoir coupling terms and Rabi field are far weaker than the photon energies. This implies that the reservoir relaxation times are fast relative to that of the system and we can safely ignore ``memory effects''~\cite{Carmichael1999}. Lastly, we make a second Markov approximation, extending the upper limit of the time integral to infinity to produce a fully Markovian equation.  This is again appropriate for a suitably rapid reservoir correlation time, requiring that the system energies are lower than the scale over which the local optical density of states (LDOS) varies~\cite{Breuer2007}. Expanding the commutator of Eq.~\eqref{eq:ME2}, %not sure if needed justification second time, just say again require rapid
\begin{gather}
\text{Tr}_R\{[  \hat{H}_{I}(t),[  \hat{H}_{I}(t-\tau), \rho_{I}R_0]]\}= \nonumber \\
\text{Tr}_R\{  \hat{H}_{I}(t)  \hat{H}_{I}(t-\tau)\rho_IR_0-  \hat{H}_{I}(t-\tau)\rho_{I}R_0  \hat{H}_{I}(t)+{\rm H.c.}\},\nonumber
\end{gather}
we now preform the trace over the reservoir, noting that each term in the above expression contains two $ \hat{H}_I$, and thus a pair of field operators. Taking the photon reservoir as a thermal bath, the only combination of field operators that will have a nonzero trace are $\text{Tr}_R\{\hat{\mathbf{f}}^\dagger (\mathbf{r}; \omega)\hat{\mathbf{f}}(\mathbf{r}'; \omega')R_0\}=n(\omega)\delta(\mathbf{r}-\mathbf{r}')\delta(\omega-\omega')$ and $\text{Tr}_R\{\hat{\mathbf{f}} (\mathbf{r}; \omega)\hat{\mathbf{f}}^\dagger(\mathbf{r}'; \omega')R_0\}= (n(\omega)+1)\delta(\mathbf{r}-\mathbf{r}') \delta(\omega-\omega')$, where the thermal photon occupation $n(\omega)=0$ for optical frequencies~\cite{Ge2013}. Thus, only one out of four components from each term survives and we have
 \begin{align}
 \dot{\rho}_{I}=&\sum_{n, n'}\int_0^\infty d\tau \int_0^\infty d\omega J_{n, n'}(\omega)e^{-i(\omega-\omega_L)\tau}\nonumber \\  &\times\left(-\hat{\sigma}^+_{n}(t)\hat{\sigma}^-_{n'}(t-\tau)\rho_{I} +\hat{\sigma}^-_{n'}(t-\tau)\rho_I\hat{\sigma}^+_n(t)\right)  +\text{H.c},\nonumber
 \end{align}
 where we used  Eq.~\eqref{eq:prop3} to evaluate 
 \begin{gather}
J_{n, n'}(\omega)=\frac{\mathbf{d}_n\cdot \text{Im}\{\mathbf{G}(\mathbf{r}_n, \mathbf{r}_{n'}; \omega)\}\cdot\mathbf{d}_{n'}}{\pi\hbar\epsilon_0} \nonumber \\
=\frac{1}{\pi\hbar\epsilon_0}\int d\mathbf{r}'' \epsilon_I(\mathbf{r}''; \omega) \mathbf{d}_n\cdot \mathbf{G}(\mathbf{r}_n, \mathbf{r}''; \omega)\cdot\mathbf{G}(\mathbf{r}'', \mathbf{r}_{n'}; \omega)\cdot\mathbf{d}_{n'} \nonumber
 \end{gather}
 the photon-reservoir spectral function, which is directly proportional to the LDOS.  From Eq.~\eqref{eq:prop3} $J_{n, n'}=J_{n', n}$, and this was used to group terms.

  We then proceed with the integration over $\tau$.  As discussed earlier $\hat{\sigma}^{\pm}(t)$ is slowly-varying; for small emitter-laser detuning and a weak Rabi field, it is appropriate to take 
$e^{\pm i(\omega-\omega_L)\tau}\sigma_{n}^\pm(t-\tau)\approx\sigma_{n}^\pm(t)e^{\pm i(\omega-\omega_L)\tau}$ as the emitter system evolves on a much slower timescale than $\omega_L$.  This also leads to the system sampling the photon LDOS at $\omega_L$, as to be expected from linear scattering theory, with, for example, a Mollow triplet centred at $\omega_L$ \cite{Carmichael1999}.  For an emitter without a laser drive however, it is more sensible to take $\sigma_{n}^\pm(t-\tau)\approx e^{-iH_S\tau/\hbar}\sigma_{n}^\pm(t)e^{iH_S\tau/\hbar}\approx \sigma_{n}^\pm(t)e^{\mp i(\omega_{n}-\omega_L)\tau}$, resulting in the operating frequency instead being $\omega_{n}$.  To keep this approach general, we take $e^{\pm i(\omega-\omega_L)\tau}\sigma_{n}^\pm(t-\tau)\approx\sigma_{n}^\pm(t)e^{\pm i(\omega-\omega_{0, n})\tau}$, with $\omega_{0, n}$ being the relevant frequency ($\omega_L$ or $\omega_{n'}$).  We note that works such as Ref.~\onlinecite{Dung2002} do not consider driven systems and avoid this complication.  For the case of a strong Rabi field, additional terms will be produced which sample the LDOS at $\omega_L\pm\frac{\Omega_R}{2}$ as well as $\omega_L$;  we refer the reader to Ref.~\onlinecite{Ge2013} for resultant master equation rate terms if this is the case.  We note that all results of the main paper were also calculated including these additional terms, and no changes were observed.  After making this approximation, preforming the integral over $\tau$, and transforming back to the Schr\"{o}dinger picture we find
  \begin{align}
 \dot{\rho}=&\frac{1}{i\hbar}[\hat{H}_{S}, \rho] \nonumber \\&+\sum_{n, n'} i\int_0^\infty d\omega \frac{J_{n, n'}(\omega)}{\omega_{0,n'}-\omega}\left(-\hat{\sigma}^+_{n}\hat{\sigma}^-_{n'}\rho +\hat{\sigma}^-_{n'}\rho\hat{\sigma}^+_n\right)  +\text{H.c}. .\nonumber
 \end{align}
 In the above, we were able to preform the reverse transformation since all operators now depend only on $t$.  In order to preform the integral over $\omega$ we note that, $J$, like $\mathbf{G}$, is analytic in the upper portion of the complex plane and use contour integration to evaluate the integral over $\omega$.  We choose a contour comprised of the real axis with an indent around the pole at $\omega=\omega_{0,n'}$ and a large semicircle in the upper complex plane, and use the relation~\cite{Riley2006, Arfken2008} $\lim_{y\to0^+}\int_{A}^B \frac{f(x)}{x+iy}dx=-i\pi\int_{A}^B f(x)\delta(x)dx+\text{P}\int_{A}^B \frac{f(x)}{x}dx$, where $B<0<A$, $\text{P}$ denotes the principal value, and $x$ is real.  It is apparent that
 \be 
i\int_0^\infty d\omega \frac{J_{n, n'}(\omega)}{\omega_{0,n'}-\omega}=\frac{\Gamma_{n, n'}}{2}-i\text{P}\int_{-\infty}^\infty \frac{J_{n,n'}(\omega)}{\omega-\omega_{0,n'}}d\omega,
\ee
where $\Gamma_{n, n'}=\frac{2}{\hbar\epsilon_0}\mathbf{d}_n\cdot\text{Im}\left\{ \mathbf{G} ( \mathbf{r}_n,   \mathbf{r}_{n'}; \omega_{0,n'})\right\}\cdot \mathbf{d}_{n'}$ and the principal value integral was extended to $-\infty$ since the principal value depends only on the relevant pole at $\omega=\omega_{0, n'}$.  We then exploit the Kramers-Kronig relations~\cite{Arfken2008}, noting that for a $f(x)$ which is analytic in the upper half plane, $\oint \frac{f(x)}{x-x_0}dx=\text{P}\int_{-\infty}^\infty \frac{f(x)}{x-x_0}dx-i\pi f(x)=0$.  Rearranging and taking the real part of both sides, it is easy to see that $\text{P}\int_{-\infty}^\infty \frac{\text{Im}\{f(x)\}}{x-x_0}dx=\pi\text{Re}\{f(x)\}$ and thus
\be
i\int_0^\infty d\omega \frac{J_{n, n'}(\omega)}{\omega_{0,n'}-\omega}=\frac{\Gamma_{n, n'}}{2}+i\delta_{n, n'}.
\label{eq:int}
\ee
As in the main text, $\delta_{n, n'}|_{n \neq n'}=\frac{-1}{\hbar\epsilon_0}\mathbf{d}_n \cdot\text{Re}\left\{ \mathbf{G} ( \mathbf{r}_n, \mathbf{r}_{n'};\omega_{0,n'})\right\}\cdot \mathbf{d}_{n'}$.  We note that in order to do this in a self-consistent fashion $\delta_{n, n}=\frac{-1}{\hbar\epsilon_0}\mathbf{d}_n \cdot\text{Re}\left\{ \mathbf{G} ( \mathbf{r}_n, \mathbf{r}_{n};\omega_{0,n})\right\}\cdot \mathbf{d}_{n}$  must be calculated first and included in the system Hamiltonian, and then the derivation must be repeated.  These terms correspond to the self-Lamb shift of each emitter, resulting in $\omega_n\to\omega_n'$ in $H_s$, $\Gamma_{n, n'}$, and $\delta_{n, n'}$.  Noting that the Hermitian conjugate term contributes $\frac{\Gamma_{n, n'}}{2}-i\delta_{n, n'}$, we find
   \begin{align}
 \dot{\rho}&=\frac{1}{i\hbar}[\hat{H}_{S}, \rho] \nonumber \\
 &+\sum_{n, n'} \frac{\Gamma_{n, n'}}{2}\left(-\hat{\sigma}^+_{n}\hat{\sigma}^-_{n'}\rho -\rho\hat{\sigma}^+_{n'}\hat{\sigma}^-_{n}+\hat{\sigma}^-_{n'}\rho\hat{\sigma}^+_n +\hat{\sigma}^-_{n}\rho\hat{\sigma}^+_{n'}\right) 
\nonumber \\
 &+i\sum_{n, n'}^{n \neq n'} \delta_{n, n'}\left(
 -\hat{\sigma}^+_{n}\hat{\sigma}^-_{n'}\rho +\rho\hat{\sigma}^+_{n'}\hat{\sigma}^-_{n}
 +\hat{\sigma}^-_{n'}\rho\hat{\sigma}^+_n -
 \hat{\sigma}^-_{n}\rho\hat{\sigma}^+_{n'}\right) .\nonumber
 \end{align}
Regrouping terms, we quickly arrive at
 \begin{align}
 \dot{\rho}=&\frac{1}{i\hbar}[\hat{H}_{S}, \rho] 
 -i\sum_{n, n'}^{n \neq n'} \delta_{n, n'}[ \hat{\sigma}^+_{n}\hat{\sigma}^-_{n'}, {\rho}] \nonumber \\
&+\sum_{n, n'}\Gamma_{n, n'}\left (\hat{\sigma}^-_{n'}{\rho}\hat{\sigma}^+_{n}-\frac{1}{2}\{\hat{\sigma}^+_{n}\hat{\sigma}^-_{n'}, {\rho}\}\right).
\end{align}

From here, it is evident that by expanding out $\hat{H}_S$ and including pure dephasing through the relevant Lindbladian term ($\mathcal{L}[\hat{O}]= (\hat{O}{\rho}\hat{O}^\dagger-\frac{1}{2}\{\hat{O}^\dagger\hat{O}, {\rho}\})$) one indeed arrives at Eq.~\eqref{eq:LME} of the main text. $\hat{H}_S=\sum_n\hbar\Delta\omega_n \hat{\sigma}^+_{n}\hat{\sigma}^-_{n}+\frac{\hbar\Omega_{R,n}}{2}(\hat{\sigma}^+_{n}+ \hat{\sigma}^-_{n})$, where $\Delta{\omega}_n =\omega_n'-\omega_L$ includes the Lamb-shift renormalized emitter resonance and is the reason those terms are omitted from the sum over $\delta_{n, n'}$. We have yet to choose $\omega_{0, n'}$ in $\gamma_{n, n'}$ and $\delta_{n, n'}$; in Eq.~\eqref{eq:LME} of the main text we have used $\omega'_{n'}$ because we initially consider an unpumped system, and then only consider resonant or near-resonant driving.  The laser detunings are either zero or far smaller than the scale over which the LDOS varies, so $\mathbf{G} ( \mathbf{r}_n, \mathbf{r}_{n'};\omega_{L})=\mathbf{G} ( \mathbf{r}_n, \mathbf{r}_{n'};\omega'_{n'})$ throughout the main document and the choice of $\omega_{0, n'}$ is unimportant. We caution however that for larger detunings $\omega_L$ should be used. Pure dephasing is introduced phenomenologically via $\sum_n\mathcal{L}[\hat{\sigma}^+_n\hat{\sigma}^-_n]$) because it can be treated to good approximation independently of the photon reservoir and depends on the nature of the emitter.  For quantum dots as considered in this work, pure dephasing through phonon interactions forms the dominant non-radiative loss mechanism~\cite{Weiler2012}.

\subsection{Derivation of the incoherent spectrum}
\label{subsec:spec}
The Heisenberg equation of motion of an operator $\hat{O}$ is found from $\dot{\hat{O}}=\frac{i}{\hbar}\left[\hat{H},\hat{O} \right]$.  Returning to Eq.~\eqref{eq:H}, it is straightforward to use the bosonic commutation relations and find for $\hat{\mathbf{f}}(\mathbf{r}; \omega)$,
\begin{align}
\dot{\hat{\mathbf{f}}} (\mathbf{r}; \omega, t)&=-i\omega\hat{\mathbf{f}} (\mathbf{r}; \omega)\nonumber \\&+  \sqrt{\frac{\epsilon_I(\mathbf{r}; \omega)}{\hbar\pi\epsilon_0}}\sum_n \mathbf{G}^*(\mathbf{r}, \mathbf{r}_n; \omega)\cdot\mathbf{d}_n\hat{\sigma}^-_n(t),
\label{eq:fmot}
\end{align}
 where we have used \eqref{eq:prop1} and made the rotating-wave approximation as was done in Sec.~\ref{sec:meq}.  From here, we take the Laplace transform of Eq.~\eqref{eq:fmot} ($\hat{O}(\omega)=\int_0^\infty\hat{O}(t)dt$):
 \begin{align}
 \hat{\mathbf{f}} (\mathbf{r}; &\omega_\lambda, \omega)=\frac{i\hat{\mathbf{f}} (\mathbf{r}; \omega_\lambda, t=0)}{\omega-\omega_\lambda}\nonumber \\&+i\sqrt{\frac{\epsilon_I(\mathbf{r}; \omega)}{\hbar\pi\epsilon_0}}\sum_n \frac{\mathbf{G}^*(\mathbf{r}, \mathbf{r}_n; \omega_\lambda)}{\omega-\omega_\lambda}\cdot\mathbf{d}_n\hat{\sigma}^-_n(\omega).
 \label{eq:flap}
 \end{align}
 In the above, we have re-indexed the field mode frequency to be $\omega_\lambda$.  In the absence of the emitter system, the field operator would evolve as $\dot{\hat{\mathbf{f}}}^0 (\mathbf{r}; \omega_\lambda, t)=-i\omega_\lambda\hat{\mathbf{f}}^0 (\mathbf{r}; \omega_\lambda, t)$ and thus $\hat{\mathbf{f}}^0 (\mathbf{r}; \omega_\lambda, \omega)=\frac{i}{\omega-\omega_\lambda}\hat{\mathbf{f}}^0 (\mathbf{r}; \omega_\lambda, t=0)$.  Noting that $\hat{\mathbf{f}}^0(t=0)=\hat{\mathbf{f}}(t=0)$, we can substitute the first term in Eq.~\eqref{eq:flap} for $\hat{\mathbf{f}}^0 (\mathbf{r}; \omega_\lambda; \omega)$.  We insert Eq.~\eqref{eq:flap} into the Laplace transformed form of Eq.~\eqref{eq:Eq} and using Eq.~\eqref{eq:prop3} find
  \begin{align}
\hat{\mathbf{E}}(\mathbf{r}; \omega_\lambda, \omega)= &  \hat{\mathbf{E}}_0(\mathbf{r}; \omega_\lambda, \omega) \label{E} \\
&- \frac {1} {\pi\epsilon_0}\sum_n  \frac{\text{Im}\left\{ \mathbf{G} ( \mathbf{r}, \mathbf{r}_n; \omega_\lambda)\right\}}{\omega-\omega_\lambda}\cdot\mathbf{d}_n \hat{\sigma}^-_n(\omega),\nonumber
\end{align}
where $\hat{\mathbf{E}}_0$ is  the  background field independent of the emitter system, defined through Eq.~\eqref{eq:EfromG} with $\hat{\mathbf{f}}^0$ instead of $\hat{\mathbf{f}}$.  To calculate the incoherent spectrum, we first need to solve for
 $\hat{\mathbf{E}}^+(\mathbf{r}; \omega)=\int_0^\infty d\omega_\lambda \hat{\mathbf{E}}(\mathbf{r}; \omega_\lambda, \omega)$.
Using the same method as was done previously to arrive at Eq.~\eqref{eq:int}:
\begin{align}
-\int_0^\infty d\omega_\lambda\frac{\text{Im}\left\{ \mathbf{G} ( \mathbf{r}, \mathbf{r}_n; \omega)\right\}}{\omega-\omega_\lambda}=i\pi\text{Im}\left\{ \mathbf{G} ( \mathbf{r}, \mathbf{r}_n; \omega)\right\}\nonumber \\
 +\pi\text{Re}\left\{ \mathbf{G} ( \mathbf{r}, \mathbf{r}_n; \omega)\right\}=\pi \mathbf{G} ( \mathbf{r}, \mathbf{r}_n; \omega), \nonumber
\end{align}
 we find
\be
\hat{\mathbf{E}}^+(\mathbf{r}; \omega)=\hat{\mathbf{E}}^+_0(\mathbf{r}; \omega) + \frac{1}{\epsilon_0}\sum_n
\mathbf{G} ( \mathbf{r}, \mathbf{r}_n; \omega)\cdot\mathbf{d}_n\hat{\sigma}^-_n(\omega).
\label{eq:Eplus}
\ee

The  detected emission spectrum at $\mathbf{r}_D$ is found by taking the Fourier transform of the first-order quantum correlation function $G^{(1)}(\mathbf{r},\tau)=\langle \hat{\mathbf{E}}^-(\mathbf{r}, t)\hat{\mathbf{E}}^+(\mathbf{r}, t+\tau)\rangle$~\cite{Carmichael1999}.  In the rotating frame the total spectrum is
\begin{align}
S^T_{D}(\omega)=&\lim_{T \to \infty}\!\frac{1}{T}\!\!\int_0^T\!\! \!\!\!\!dt\!\!\int_0^T \!\!\!\!\!\!dt'\!
 \langle\hat{\mathbf{E}}^- (\mathbf{r}_D, t) \hat{\mathbf{E}}^+(\mathbf{r}_D, t')\rangle e^{ i(\omega_L - \omega ) (t-t') }\nonumber \\
=&\lim_{T \to \infty}\frac{1}{T}\langle\hat{\mathbf{E}}^- (\mathbf{r}_D; \omega) \hat{\mathbf{E}}^+(\mathbf{r}_D; \omega)\rangle.
\end{align}
 Inserting Eq.~\eqref{eq:Eplus} and its Hermitian conjugate into the above we find
\be
\langle\hat{\mathbf{E}}^- (\mathbf{r}_D; \omega) \hat{\mathbf{E}}^+(\mathbf{r}_D; \omega)\rangle=\sum_{n, n'}g_{n, n'}(\omega)\langle\hat{\sigma}_n^+(\omega) \hat{\sigma}_{n'}^-(\omega)\rangle
\label{eq:epem}
\ee
where the emitter coupling term
$g_{n,n'}(\omega)=\frac{1}{\epsilon_0^2}\mathbf{d}_n\cdot\mathbf{G}^*(\mathbf{r}_{n}, \mathbf{r}_D; \omega)\cdot\mathbf{G}(\mathbf{r}_D, \mathbf{r}_{n'}; \omega)\cdot\mathbf{d}_{n'}$ using Eq.~\eqref{eq:prop1} and it is apparent that $g_{n',n}=g^*_{n, n'}$. As discussed earlier thermal effects are negligible at optical frequencies, and so we have taken the free field to be in the vacuum state, eliminating the terms containing  $\hat{\mathbf{E}}^\pm_0$ in  Eq.~\eqref{eq:epem}~\cite{Carmichael1999}.  Since vacuum free field and the rotating-wave approximations were made in Sec.~\ref{sec:meq}, it is important to also make them here so that we calculate the emitted spectrum of the actual system considered.

  We would like to write the spectrum as a convolution of atomic operators in the time domain, and so expand
\begin{align}
\lim_{T \to \infty}\frac{1}{T}\langle\hat{\sigma}_n^+(\omega) \hat{\sigma}_{n'}^-(\omega)\rangle=&\lim_{T \to \infty}\frac{1}{T}\int_0^T dt\left(\int_0^t dt'+ \int_t^T dt'\right)\nonumber \\
&\times\langle\hat{\sigma}_n^+(t) \hat{\sigma}_{n'}^-(t')\rangle e^{ i(\omega_L - \omega ) (t-t')}.\nonumber
\end{align}
we then define $\tau=t-t'$ in the first integral, and $\tau=t'-t$ in the second such that it remains a positive quantity.  We find
\begin{align}
\lim_{T \to \infty}\frac{1}{T}&\langle\hat{\sigma}_n^+(\omega) \hat{\sigma}_{n'}^-(\omega)\rangle=\lim_{T \to \infty}\frac{1}{T}\int_0^T dt\nonumber \\\times\Bigg(&
\int_0^t \!\!\!\!d\tau \langle\hat{\sigma}_n^+(t+\tau) \hat{\sigma}_{n'}^-(t)\rangle e^{ i(\omega_L - \omega )\tau} \nonumber \\&~~~~~~~~~~+ \int_0^{T-t}\!\! \!\!\!\! \!\! \!\!\!\!d\tau\langle\hat{\sigma}_n^+(t) \hat{\sigma}_{n'}^-(t+\tau)\rangle e^{- i(\omega_L - \omega )\tau}\Bigg). \nonumber
\end{align}
In the above, we used the fact that $\langle\hat{\sigma}^+(t+\tau) \hat{\sigma}^-(t)\rangle$ depends only on the separation $\tau$ at which each operator is evaluated~\cite{Carmichael1999} to take $\langle\hat{\sigma}_n^+(t) \hat{\sigma}_{n'}^-(t-\tau)\rangle=\langle\hat{\sigma}_n^+(t+\tau) \hat{\sigma}_{n'}^-(t)\rangle$.  Since these terms thus have no $t$-dependence  we are free to preform the outermost integral and extend $T$ to infinity~\cite{Meystre1999}:
\begin{align}
\lim_{T \to \infty}\frac{1}{T}\langle\hat{\sigma}_n^+(\omega) \hat{\sigma}_{n'}^-(\omega)\rangle=&\lim_{t \to \infty} \!\!\int_0^\infty \!\! \!\!\!\!d\tau \langle\hat{\sigma}_n^+(t+\tau) \hat{\sigma}_{n'}^-(t)\rangle e^{ i(\omega_L - \omega )\tau}\nonumber \\ &+ \int_0^{\infty} \! \!\! \!\!\!\!d\tau\langle\hat{\sigma}_n^+(t) \hat{\sigma}_{n'}^-(t+\tau)\rangle e^{- i(\omega_L - \omega )\tau}.\nonumber
\end{align}
 Noting that the two integrals are Hermitian conjugates and inserting this into Eq.~\eqref{eq:epem} we arrive at an expression for the total emitted spectrum in terms of readily solvable quantities using the master equation approach:
\begin{align}
&S^T_{D}(\omega)=\sum_{n, n'}2g_{n, n'}(\omega)\nonumber \\ &\times
\text{Re}\{\lim_{t \to \infty}\int_0^\infty d\tau \langle\hat{\sigma}_n^+(t+\tau) \hat{\sigma}_{n'}^-(t)\rangle e^{ i(\omega_L - \omega )\tau}\}
\label{eq:totspec}
\end{align}

 Of particular interest is not the total emitted spectrum but the incoherent spectrum, which contains the quantum dynamics of the system~\cite{Carmichael1999}.   This is found by subtracting from the total spectrum the coherent portion, defined as
\begin{align}
S^{\rm coh}_{D}(\omega)=&\lim_{t \to \infty}\int_{-\infty}^\infty d\tau|\langle\hat{\mathbf{E}}^+(\mathbf{r}_D, t)\rangle|^2e^{- i(\omega_L - \omega ) \tau }\nonumber \\=&|\langle\hat{\mathbf{E}}_{ss}^+(\mathbf{r}_D)\rangle|^2(\omega)
\end{align}
where $\hat{\mathbf{E}}_{ss}$ denotes the steady state ($t\to \infty$) value~\cite{Carmichael1999} and the integral is extended to $-\infty$ to include the Hermitian conjugate term.  We note that $|\langle\hat{\mathbf{E}}_{ss}^+(\mathbf{r}_D)\rangle|^2(\omega)=2\pi|\langle\hat{\mathbf{E}}_{ss}^+(\mathbf{r}_D)\rangle|^2\delta(\omega-\omega_L)$ and this term thus produces a Dirac delta peak in the emitted spectra at the laser frequency, as expected for energy-conserving coherent scattering~\cite{Meystre1999}.  Again dropping the free field terms and writing this in terms of atomic operators  we quickly find
\begin{align}
&S^{\rm coh}_{D}(\omega)=\sum_{n, n'}g_{n, n'}(\omega)\langle\hat{\sigma}_{ss,n}^+\rangle \langle\hat{\sigma}_{ss,n'}^-\rangle(\omega)\label{eq:cospec} \\ &=\sum_{n, n'}g_{n, n'}(\omega)\lim_{t\to\infty}\int_{-\infty}^\infty \!\! \!\!\!\!d\tau \langle\hat{\sigma}_{n}^+(t)\rangle \langle\hat{\sigma}_{n'}^+(t)\rangle e^{ -i(\omega_L - \omega ) \tau },\nonumber \\
&=\sum_{n, n'}2g_{n, n'}(\omega)\text{Re}\{\lim_{t\to\infty}\int_{0}^\infty \!\! \!\!\!\!d\tau \langle\hat{\sigma}_{n}^+(t)\rangle \langle\hat{\sigma}_{n'}^+(t)\rangle e^{ i(\omega_L - \omega ) \tau }\}. \nonumber
\end{align}
This is then simply subtracted from  Eq.~\eqref{eq:totspec} to find the incoherent spectrum.  As in the main text, we drop extraneous numerical factors and separate this into direct and interference terms
\begin{align}
S_{D}(\omega)=&\sum_n|\mathbf{G}(\mathbf{r}_D, \mathbf{r}_n; \omega)\cdot\frac{\mathbf{d}_n}{\epsilon_0}|^2
 {\rm Re}\{ S^0_{ n, n}(\omega)\} \nonumber \\
&+\sum_{n, n'}^{n\neq n'}\text{Re}\{g_{n, n'}(\omega)S^0_{n, n'}(\omega)\},
\label{eq:inspec}
\end{align}

where we have defined a bare incoherent spectrum,
\begin{align}
 S^0_{n, n'}(\omega)=&\lim_{t \to \infty} \int_0^\infty\!\! \!\!\!\! d\tau ( \langle\hat{\sigma}^+_n (t+\tau) \hat{\sigma}^-_{n'}(t)\rangle-\langle\hat{ \sigma }^+_n(t)\rangle\langle\hat{\sigma}^-_{n'}(t)\rangle )\nonumber \\
 &\times e^{ i(\omega_L - \omega ) \tau }.
\end{align}
 The first sum in Eq.~\eqref{eq:inspec} corresponds to the incoherent spectrum emitted from a single emitter which is measured by a detector, although the effect of the surrounding emitters is still seen in the $ S^0_{ n, n}$ term due to its influence on the expectation value $\langle\hat{\sigma}^+_n (t+\tau) \hat{\sigma}^-_{n'}(t)\rangle$.  The second term is due to quantum interference and will be zero for a system containing a single emitter.  It includes ``which-path'' information, describing light from a single emitter scattering off another before propagating to the detector.
\vspace{100pt}

%We thank  Rong-Chun Ge for assistance with computational software.
%\end{acknowledgements}
%\bibliographystyle{osajnl}
\bibliography{paperquantbib}
\end{document}